\documentclass[11pt]{article}
\usepackage{graphicx}
\usepackage{cite}
\usepackage{amssymb}
\usepackage{epstopdf}
\usepackage{hep}
\usepackage{feynmp}
\usepackage{multicol}
\usepackage{simplewick}
\usepackage{appendix}
\usepackage{subfigure}
\usepackage{color}
\usepackage{ulem}
\newcommand{\fig}[2]{\scalebox{#1}{\includegraphics{#2}}}

\newcommand{\vb}{\biggl|}

\newcommand{\Sh}{\hat{S}}
\newcommand{\Tr}[1]{{\rm Tr}\left[ #1 \right]}
\newcommand{\cl}{\biggr|_{\rm c.l.}}

\newcommand{\nn}{\nonumber}

\allowdisplaybreaks

\begin{document}
\title{{\bf Single-spin asymmetries in the leptoproduction of transversely polarized {\boldmath $\Lambda$} hyperons}}

\author{K.~Kanazawa$^{1}$, A.~Metz$^{1}$, D.~Pitonyak$^{2}$, and M.~Schlegel$^{3}$
 \\[0.3cm]
{\normalsize\it $^1$Department of Physics, SERC,
  Temple University, Philadelphia, PA 19122, USA} \\[0.15cm]
{\normalsize\it $^2$RIKEN BNL Research Center,
                 Brookhaven National Laboratory,
                 Upton, New York 11973, USA} \\[0.15cm]
{\normalsize\it $^3$Institute for Theoretical Physics, T\"ubingen University,} \\ 
{\normalsize\it Auf der Morgenstelle 14, D-72076 T\"ubingen, Germany}            
}

\date{\today}
\maketitle

\begin{abstract}
\noindent
We analyze single-spin asymmetries (SSAs) in the leptoproduction of transversely polarized $\Lambda$ hyperons within the collinear twist-3 formalism.  We calculate both the distribution and fragmentation terms in two different gauges (lightcone and Feynman) and show that the results are identical.  This is the first time that the fragmentation piece has been analyzed for transversely polarized hadron production within the collinear twist-3 framework.  In lightcone gauge we use the same techniques that were employed in computing the analogous piece in $p^\uparrow p\to \pi\,X$, which has become an important part to that reaction.  With this in mind, we also verify the gauge invariance of the formulas for the transverse SSA in the leptoproduction of pions.
\end{abstract}

%
%
\section{Introduction} \label{s:Intro}
The first measurement of transverse single-spin asymmetries (SSAs), denoted $A_N$, back in the 1970s was in polarized $\Lambda$ production from proton-beryllium collisions, which revealed quite large effects~\cite{Bunce:1976yb}.  After the na\"{i}ve collinear parton model failed to generate these large asymmetries~\cite{Kane:1978nd} (see also \cite{Dharmaratna:1989jr}), Efremov and Teryaev realized that one must go beyond this framework and also include quark-gluon-quark correlations in the nucleon~\cite{Efremov:1981sh}.  This formalism, known as collinear twist-3 factorization, was first worked out in detail by Qiu and Sterman and applied to reactions where one of the incoming nucleons is transversely polarized~\cite{Qiu:1991pp,Qiu:1998ia} instead of the final-state hadron.  During the same time, another mechanism, known as the Generalized Parton Model (GPM), was also put forth to explain transverse SSAs in proton-proton collisions~\cite{Anselmino:1994tv,Anselmino:1998yz}.  This approach involves the Sivers~\cite{Sivers:1989cc} and Collins~\cite{Collins:1992kk} transverse momentum dependent (TMD) functions.  Over the last several decades, processes  like $p^\uparrow p\to C\,X$, where $C$ is a light, unpolarized hadron or jet, have been the subject of much intense theoretical~\cite{Efremov:1981sh,Qiu:1991pp,Qiu:1998ia,Kouvaris:2006zy,Kang:2011hk,Kanazawa:2000hz,Kanazawa:2000kp,Koike:2009ge,Kanazawa:2010au,Kanazawa:2011bg,Beppu:2013uda,Kang:2010zzb,Metz:2012ct,Kanazawa:2014dca,Anselmino:1994tv,Anselmino:1998yz,Anselmino:2005sh,Anselmino:2012rq,Anselmino:2013rya} and experimental~\cite{Adams:1991rw,Krueger:1998hz,Adams:2003fx,Adler:2005in,Lee:2007zzh,:2008mi,Adamczyk:2012qj,Adamczyk:2012xd,Bland:2013pkt,Adare:2013ekj} work.  

Most recently, within the collinear twist-3 approach, the fragmentation mechanism has been put forth as the main cause of $A_N$ in $p^\uparrow p\to \pi\,X$~\cite{Kanazawa:2014dca}.  This comes after many years of assuming the so-called Qiu-Sterman (QS) function is the origin of this asymmetry~\cite{Qiu:1998ia,Kouvaris:2006zy}, which led to the infamous ``sign mismatch'' between the QS function and the Sivers function~\cite{Kang:2011hk}.  Given the potential significant role of the fragmentation term in $p^\uparrow p\to \pi\,X$, which was first fully derived in~\cite{Metz:2012ct}\footnote{The so-called ``derivative term'' was first computed in Ref.~\cite{Kang:2010zzb}.} by two of the authors (A.M.~and D.P.) in lightcone gauge, it is important to see if the techniques employed there lead to consistent, gauge invariant results for other SSA processes.\footnote{The fragmentation term has also been calculated in the collinear twist-3 approach for asymmetries in $\ell \,p^\uparrow\to \ell^\prime\pi\,X$~\cite{Yuan:2009dw,Kanazawa:2013uia}, $\ell \,p^\uparrow\to \pi\,X$~\cite{Gamberg:2014eia}, and $\vec{\ell} \,p^\uparrow\!\to \pi\,X$~\cite{Kanazawa:2014tda}.}  To this end, we analyze the transverse SSA in polarized $\Lambda$ production from lepton-proton collisions, i.e., $\ell\,p\to \Lambda^\uparrow X$, in both lightcone gauge and Feynman gauge, where for the former we follow the procedure in Ref.~\cite{Metz:2012ct} for the fragmentation piece.  These are the novel results from this work.  In addition, we verify that the formulas in~\cite{Gamberg:2014eia} for $\ell\,p^\uparrow\to \pi\, X$, especially the fragmentation term, are gauge invariant.  We mention that, although given less attention, transverse SSAs in polarized hyperon production have been explored before, both in the collinear twist-3~\cite{Kanazawa:2000cx,Zhou:2008fb} and GPM~\cite{Anselmino:2000vs} frameworks.  However, this will be the first time that the fragmentation piece has been analyzed in the collinear twist-3 formalism for transversely polarized hadron production.

The paper is organized as follows: first, in Sec.~\ref{s:Def} we define the relevant non-perturbative functions.  Next, in Sec.~\ref{s:LC} we give a few details of the derivation in lightcone gauge of the single-spin dependent cross section for $\ell\,p\to \Lambda^\uparrow X$.  This includes both the distribution and fragmentation terms.  Then in Sec.~\ref{s:Feyn} we repeat our computation but in Feynman gauge.  Finally, in Sec.~\ref{s:sum} we summarize our results and conclude the paper.

%
%
\section{Definitions of the non-perturbative functions} \label{s:Def}
In this section we define the relevant non-perturbative functions for our computation of the leading-order (LO) single-spin dependent cross section for the process
\begin{equation}
\ell(l) + p(P) \rightarrow \Lambda(P_h,S_{hT}) + X\,, \label{e:ANreac}
\end{equation}
where the momenta and polarizations of the particles are given.  Note that the transverse spin of the $\Lambda$ hyperon $\vec{S}_{hT}$ is understood as being w.r.t.~its momentum $\vec{P}_h$.  We mention that although tri-gluon fragmentation functions (FFs) are nonzero for a transversely polarized hadron, they will not be relevant for our LO calculation, so we will not discuss them here.

For the fragmentation term we need the chiral-even collinear twist-3 fragmentation correlators for a transversely polarized spin-1/2 hadron, which will enter our result coupled to the twist-2 unpolarized parton distribution function (PDF) $f_1(x)$, defined in the usual way~\cite{Collins:1981uw,Collins:book}.  First, for the TMD twist-2 quark-quark correlator\footnote{These twist-2 TMD FFs will give us collinear twist-3 FFs through their first transverse-momentum moments (see Eqs.~(\ref{e:Dhat}), (\ref{e:Ghat})).} we have~\cite{Boer:1997mf}
\begin{align}
\sum_{X}\hspace{-0.55cm}\int\int\! &\frac{d (n_h\cdot\xi) d^2\vec{\xi}_T} {(2\pi)^3}e^{ik\cdot \xi}\langle 0|\psi^q_i(\xi)|P_h,S_{hT};X\rangle\langle P_h,S_{hT};X|\bar{\psi}^q_j(0)|0\rangle \bigg |_{\bar{n}_h\cdot\xi = \,0}\nonumber\\[0.2cm]
&=\frac{z} {M_h}\epsilon_T^{k_T S_{hT}}(\slashed{n}_{h})_{ij}\,D_{1T}^{\perp,q}(z,z^2\vec{k}_T^2) +\frac{z} {M_h}k_T\cdot S_{hT} (\slashed{n}_{h}\gamma_5)_{ij}\, G_{1T}^q(z,z^2\vec{k}_T^2)\,.\label{e:DeltaTCorrel}
\end{align}
Note that we will suppress all Wilson lines unless they are pertinent to our discussion.  In Eq.~(\ref{e:DeltaTCorrel}), we have $\epsilon_T^{\alpha\beta} \equiv \epsilon^{\bar{n}_h n_h \alpha\beta} \,\,(\epsilon^{0123} = +1)$, where $n_h\sim P_h$ with $n_h\cdot \bar{n}_h = 1$.  Next, for the genuine twist-3 collinear quark-quark FFs, we have
\begin{align}
z^2\sum_{X}\hspace{-0.55cm}\int\int\! &\frac{d (n_h\cdot\xi) } {2\pi}e^{ik\cdot \xi}\langle 0|\psi^q_i(\xi)|P_h,S_{hT};X\rangle\langle P_h,S_{hT};X|\bar{\psi}^q_j(0)|0\rangle \bigg |_{\bar{n}_h\cdot\xi \,=\, \vec{\xi}_T = \,0}\nonumber\\[0.2cm]
&=\frac{z M_h} {\bar{n}_h\cdot P_h}\epsilon_T^{\alpha S_{hT}}(\gamma_{\alpha})_{ij}\,D_T^q(z)-\frac{z M_h} {\bar{n}_h\cdot P_h}S_{hT}^\alpha(\gamma_\alpha\gamma_5)_{ij}\,G_T^q(z)\,.  \label{e:DeltaT3}
\end{align}
We remark that the functions $G_{1T}^q(z,z^2\vec{k}_T^2)$ and $G_T^q(z)$ will not contribute to this process because their hard factors vanish (but they would be nonzero for the double-spin asymmetry $A_{LT}$ for a longitudinally polarized lepton or proton and a transversely polarized $\Lambda$ hyperon).  Finally, the so-called F-type and D-type quark-gluon-quark correlators give, respectively,
\begin{align}
\sum_{X}\hspace{-0.55cm} \int\, \!\int\! &\frac{d(n_h\cdot\xi)} {2\pi}\!\int\! \frac{d(n_h\cdot\zeta)} {2\pi} e^{i\frac{(\bar{n}_h\cdot P_{h})} {z_{1}}(n_h\cdot\xi)} e^{i\left(\frac{1} {z}-\frac{1} {z_1}\right)(\bar{n}_h\cdot P_{h})(n_h\cdot\zeta)} \nonumber\\[-0.15cm]
&\hspace{2.5cm}\times\,\langle 0|igF_{T}^{\bar{n}_h\alpha}(n_h\cdot\zeta)\psi^q_{i}(n_h\cdot\xi)|P_{h},S_{hT};X\rangle\langle P_{h},S_{hT};X|\bar{\psi}^q_{j}(0)|0\rangle\nonumber\\[0.2cm]
&=-izM_h\epsilon_T^{\alpha S_{hT}}(\slashed{n}_h)_{ij}\,\hat{D}_{FT}^q(z,z_1) - zM_h S_{hT}^\alpha(\slashed{n}_h\gamma_5)_{ij}\,\hat{G}_{FT}^q(z,z_1)\,, \label{e:F-typeFFC} \\[0.5cm]
\sum_{X}\hspace{-0.55cm} \int\, \!\int\! &\frac{d(n_h\cdot\xi)} {2\pi}\!\int\! \frac{d(n_h\cdot\zeta)} {2\pi} e^{i\frac{(\bar{n}_h\cdot P_{h})} {z_{1}}(n_h\cdot\xi)} e^{i\left(\frac{1} {z}-\frac{1} {z_1}\right)(\bar{n}_h\cdot P_{h})(n_h\cdot\zeta)} \nonumber\\[-0.15cm]
&\hspace{2.5cm}\times\,\langle 0|iD_{T}^{\alpha}(n_h\cdot\zeta)\psi^q_{i}(n_h\cdot\xi)|P_{h},S_{hT};X\rangle\langle P_{h},S_{hT};X|\bar{\psi}^q_{j}(0)|0\rangle\nonumber\\[0.2cm]
&=-\frac{izM_h} {\bar{n}_h\cdot P_h}\epsilon_T^{\alpha S_{hT}}(\slashed{n}_h)_{ij}\,\hat{D}_{DT}^q(z,z_1) - \frac{zM_h} {\bar{n}_h\cdot P_h} S_{hT}^\alpha(\slashed{n}_h\gamma_5)_{ij}\,\hat{G}_{DT}^q(z,z_1)\,. \label{e:D-typeFFC}
\end{align}
We emphasize that the quark-gluon-quark FFs have both real and imaginary parts, unlike the quark-quark ones that are purely real, and we denote these, respectively, by $\Re$ and $\Im$ superscripts.  One can establish the following relations between the F- and D-type functions:
\begin{align}
\hat{D}_{DT}^q(z,z_1) &= -\frac{i} {z^2}\,\hat{D}^q_T(z)\,\delta(1/z-1/z_1) + PV\frac{1} {1/z-1/z_1}\,\hat{D}^q_{FT}(z,z_1)\,,\label{e:FDrel1}\\[0.1cm]
\hat{G}_{DT}^q(z,z_1) &= \frac{1} {z^2}\,\hat{G}^q_T(z)\,\delta(1/z-1/z_1) + PV\frac{1} {1/z-1/z_1}\,\hat{G}^q_{FT}(z,z_1)\,, \label{e:FDrel2}
\end{align}
where 
\begin{align}
\hat{D}^q_T(z) &= z^2\int \!d^2\vec{k}_T \frac{\vec{k}_T^2} {2M_h^2}\, D_{1T}^{q,\perp}(z,z^2\vec{k}_T^2)\,,\label{e:Dhat}\\[0.1cm]
\hat{G}^q_T(z) &= z^2\int \!d^2\vec{k}_T \frac{\vec{k}_T^2} {2M_h^2}\, G_{1T}^{q}(z,z^2\vec{k}_T^2)\,. \label{e:Ghat}
\end{align}
In (\ref{e:FDrel1}), (\ref{e:FDrel2}) $PV$ stands for principal value.  One also has the following QCD equation of motion (EOM) relation:
\begin{equation}
\int_z^\infty\! \frac{dz_1} {z_1^2}\!\left[\hat{D}^q_{DT}(z,z_1)-\hat{G}^q_{DT}(z,z_1)\right] = \frac{1} {z^3}\left(iD_T^q(z)-G_T^q(z)\right),
\end{equation}
which when combined with Eqs.~(\ref{e:FDrel1}), (\ref{e:FDrel2}) leads to the useful formula
\begin{equation}
G_T^q(z) - iD_T^q(z) = z\!\left(i\hat{D}_T^q(z) + \hat{G}^q_T(z)\right) + z^3\!\int_z^\infty\!\frac{dz_1} {z_1^2}\frac{1} {1/z-1/z_1}\left[\hat{G}_{FT}^{q}(z,z_1)-\hat{D}_{FT}^{q}(z,z_1)\right]\,, \label{e:useful}
\end{equation}
where we have dropped the $PV$ in the last term because the integration does not cross over the pole at $z=z_1$.

For the distribution term we will need the twist-2 transversity FF $H^q_1(z)$ for a transversely polarized spin-1/2 hadron, which is given by~\cite{Boer:1997mf}
\begin{align}
z^2\sum_{X}\hspace{-0.55cm}\int\int\! &\frac{d (n_h\cdot\xi) } {2\pi}e^{ik\cdot \xi}\langle 0|\psi^q_i(\xi)|P_h,S_{hT};X\rangle\langle P_h,S_{hT};X|\bar{\psi}^q_j(0)|0\rangle \bigg |_{\bar{n}_h\cdot\xi \,=\,\vec{\xi}_T = \,0}= z\, \big(\slashed{S}_{hT} \slashed{n}_{h}\gamma_5\big)_{ij}\,H_1(z)\,.  \label{e:DeltaT2}
\end{align}
This FF couples to the (F-type) unpolarized chiral-odd collinear twist-3 function $H^q_{FU}(x,x_1)$, defined as
\begin{align}
\int\frac{d(\bar{n}\cdot \xi)} {2\pi}&\int\frac{d(\bar{n}\cdot \zeta)} {2\pi}\,e^{i(x_1n\cdot P)(\bar{n}\cdot \xi)}e^{i((x-x_1)n\cdot P)(\bar{n}\cdot\zeta)} \langle P|\bar{\psi}^q_j(0)igF_\perp^{n\alpha}(\bar{n}\cdot\zeta)\psi^q_i(\bar{n}\cdot\xi)|P\rangle  \nonumber\\[0.1cm]
&= -\frac{M} {2}\,i\epsilon_\perp^{\alpha\beta}\big(\gamma_\beta\slashed{\bar{n}}\gamma_5\big)_{ij}\,H_{FU}^q(x,x_1)\,, \label{e:HFU}
\end{align}
where $\epsilon_\perp^{\alpha\beta} \equiv \epsilon^{\bar{n} n \alpha\beta}$ with $\bar{n}\sim P$ and $n\cdot \bar{n} = 1$.  We will find in our calculation that $H^q_{FU}(x,x_1)$ will be evaluated at the soft-gluon pole (SGP) $x=x_1$.  It turns out $H^q_{FU}(x,x)$ has a model-independent relation to the first $p_\perp$-moment of the Boer-Mulders function $h_1^{\perp q}(x,\vec{p}_\perp^{\,2})$~\cite{Boer:2003cm},
\begin{equation}
\int \!d^2\vec{p}_\perp\,\frac{\vec{p}_\perp^{\,2}} {2M^2}\,h_1^{\perp q}(x,\vec{p}_\perp^{\,2})\big |_{\rm SIDIS} = \pi\,H_{FU}^q(x,x)\,, \label{e:HFU_BM}
\end{equation}
where~\cite{Boer:1997nt,Goeke:2005hb}
\begin{equation}
 \int\!\frac{d (\bar{n}\cdot\xi) d^2\vec{\xi}_T} {(2\pi)^3}e^{ip\cdot \xi}\langle P|\bar{\psi}^q_j(0)\mathcal{W}(0,\xi)\psi^q_i(\xi)|P\rangle \bigg |_{n\cdot\xi = \,0} = \frac{1}{2M} \sigma^{\alpha\nu} p_{\perp\alpha}\,\bar{n}_\nu\,h_1^{\perp q}(x,\vec{p}_\perp^{\,2})\,,
\end{equation}
with the Wilson line $\mathcal{W}(0,\xi)$ chosen to be consistent with the SIDIS process.

%
%
\section{Lightcone gauge calculation} \label{s:LC}
\begin{figure}[t]
 \begin{center}
 \fig{0.5}{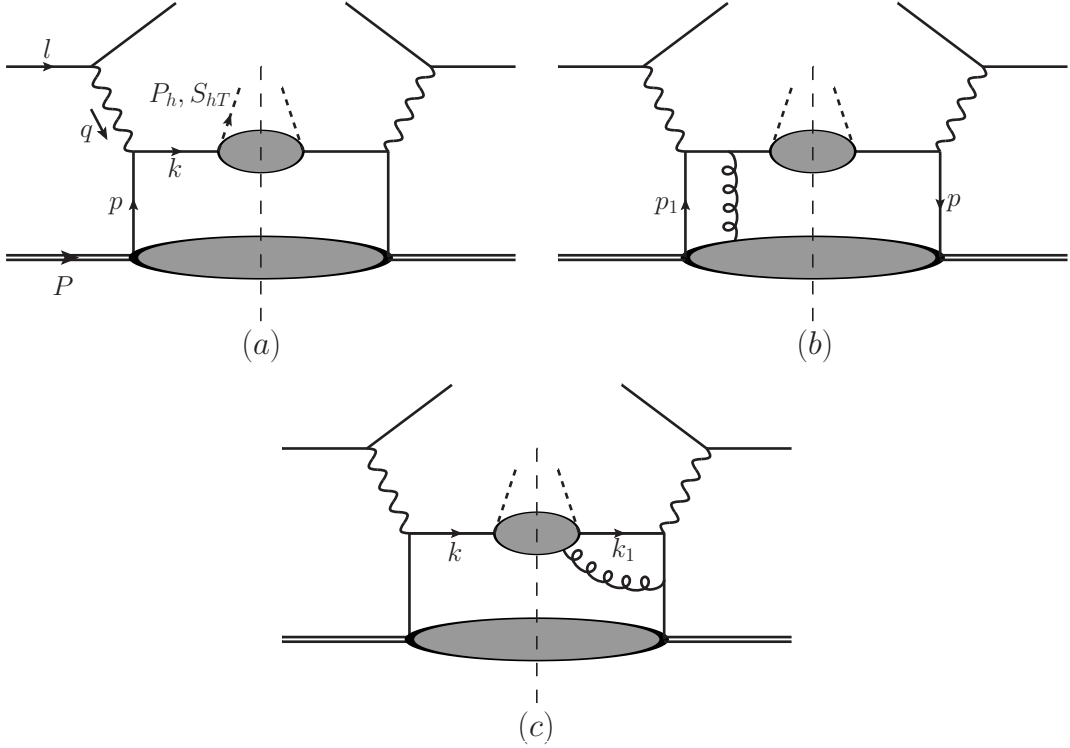}
 \end{center}
 \vspace{-0.75cm}
\caption{Graphs showing the relevant channels for $\ell\, p\to\Lambda^\uparrow X$ within collinear twist-3 factorization.  The diagram in (a) involves $qq$ correlators, while the ones in (b),$\,$(c) (and their Hermitian conjugates) deal with $qgq$ correlators.}
\label{f:graphs}
\end{figure}

In this section we outline some of the steps for the computation of the single-spin dependent cross section for $\ell\,p\to\Lambda^\uparrow X$ in lightcone gauge, where the relevant diagrams are shown in Fig.~\ref{f:graphs}.  Recall that in lightcone gauge one has $A^+ = 0$, where $+$ here is used to indicate the ``large'' component of the gluon field in the quark-gluon-quark correlators.  That is, we use $\bar{n}_h\cdot A = 0$ for the fragmentation term and $n\cdot A = 0$ for the distribution term.  However, this constraint does not completely fix the gauge, as one must also impose a boundary condition (BC) on the transverse component of the gauge field at lightcone infinity (see, e.g., \cite{Belitsky:2002sm, Boer:1997bw}).  For the fragmentation term we choose the antisymmetric BC $A_T(+\infty) + A_T(-\infty) = 0$, while for the distribution term we pick the advanced BC $A_\perp(+\infty) = 0$.  We will work in the lepton-nucleon center-of-mass frame with the nucleon moving along the $+z$-axis and the transverse momentum of the outgoing hadron, $\vec{P}_{h\perp}$, along the $+x$-axis.  The Mandelstam variables for the process are defined as $S = (P+l)^{2}$, $T = (P-P_h)^{2}$, and $U = (l-P_h)^{2}$, which on the partonic level give $\hat{s} = xS$, $\hat{t} = xT/z$, and $\hat{u} = U/z$.

We start with the calculation of the fragmentation term.  The procedure for obtaining this piece in lightcone gauge was already laid out in Ref.~\cite{Metz:2012ct}.  Therefore, we only highlight the major steps of the derivation.  First we look at the contribution from the graph in Fig.~\ref{f:graphs}(a).  In this case we keep the transverse momentum $k_T$ of the fragmenting quark, leading to a ``derivative'' and ``non-derivative'' term involving $\hat{D}_T^q(z)$, and then we neglect $k_T$ and use a twist-3 Dirac projection to generate a term involving $D_T^q(z)$. This leads to the following expression for the quark-quark ($qq$) correlator term:
\begin{align}
\frac{P_h^0d\sigma^{Frag}_{LC,qq}} {d^3\vec{P}_h} &= \frac{8M_h\alpha_{em}^2} {S}\,\epsilon_\perp^{P_{h\perp}S_{hT}}\sum_q e_q^2\int_{z_{min}}^1\!\frac{dz} {xz^3}\frac{1} {S+T/z}\,\frac{1} {-\hat{t}-x\hat{u}}\,f_1^q(x)\nonumber\\
&\hspace{0.5cm}\times\left\{z\frac{d\hat{D}_T^q(z)} {dz}\!\left[\frac{(1-x)(\hat{s}^2+\hat{u}^2)} {2\hat{t}^2}\right] - \hat{D}_T^q(z)\left[\frac{\hat{s}(\hat{t}^2(x+1)+\hat{t}\hat{u}(3x+2)+4\hat{u}^2x)} {2\hat{t}^3}\right]\right.\nonumber\\
&\hspace{1.25cm}\left. +\,\frac{1} {z}\,D_T^q(z)\left[\frac{\hat{s}(\hat{u}-(1-x)\hat{s})} {2\hat{t}^2}\right]\right\}\,, \label{e:qqLC}
\end{align}
where $z_{min} = -(T+U)/S$ and $x = -(U/z)/(S+T/z)$.  We use $d\sigma^{Frag}_{LC}$ to indicate the fragmentation term in lightcone gauge.

Next, we look at the contribution from the graph in Fig.~\ref{f:graphs}(c).  In lightcone gauge we neglect the transverse momenta of the quarks and attach transversely polarized gluon fields $A_T$.  This initially leads to a matrix element involving $A_T$, which we then ``invert'' through partial integration to obtain a gauge invariant correlator involving $F_{T}^{\bar{n}_h\alpha}$ (like in Eq.~(\ref{e:F-typeFFC})).  As a result of this procedure, we are also left with a factor of $1/(1/z-1/z_1)$ and must regulate the pole at $z=z_1$.  This prescription is determined by the BC~\cite{Belitsky:2002sm, Boer:1997bw}, which for the antisymmetric one is $1/(1/z-1/z_1)\to PV[1/(1/z-1/z_1)]$. This is a convenient choice since our result will not pick out the poles of the fragmentation correlators because 3-parton F-type FFs vanish at all partonic poles~\cite{Gamberg:2008yt,Meissner:2008yf,Gamberg:2010uw}, a property that is intimately connected to the universality of TMD FFs~\cite{Metz:2002iz,Collins:2004nx,Yuan:2007nd,Yuan:2008yv,Yuan:2009dw}. 

However, we still need to generate an imaginary phase from this diagram, which is achieved by using the non-pole pieces of $\hat{D}_{FT}^{q,\Im}(z,z_1)$, $\hat{G}_{FT}^{q,\Im}(z,z_1)$.  In the end we find the following formula for the quark-gluon-quark ($qgq$) correlator term:
\begin{align}
&\frac{P_h^0d\sigma^{Frag}_{LC,qgq}} {d^3\vec{P}_h} = \frac{8M_h\alpha_{em}^2} {S}\,\epsilon_\perp^{P_{h\perp}S_{hT}}\sum_q e_q^2\int_{z_{min}}^1\!\frac{dz} {xz}\frac{1} {S+T/z}\,f_1^q(x)\nonumber\\
&\times \int_z^\infty\!\frac{dz_1} {z_1^2} \,\frac{1} {1/z-1/z_1}\left\{\!\left(\hat{G}_{FT}^{q,\Im}(z,z_1) - \hat{D}_{FT}^{q,\Im}(z,z_1)\right)\!\left[\frac{\hat{s}(\hat{u}-\hat{s})} {2\hat{t}^3}\right] -\hat{D}_{FT}^{q,\Im}(z,z_1) \!\left[\frac{\hat{s}x(\hat{s}^2+\hat{u}^2)} {\xi\hat{t}^3(-\hat{t}-x\hat{u})}\right]\right\}\,, \label{e:qgqLC}
\end{align}
where $\xi = 1-z/z_1$.  Combining Eqs.~(\ref{e:qqLC}), (\ref{e:qgqLC}) and using the relation (\ref{e:useful}) to simplify the first term in Eq.~(\ref{e:qgqLC}), we obtain the fragmentation piece for $\ell\,p\to\Lambda^\uparrow X$ in lightcone gauge:
\begin{align}
\frac{P_h^0d\sigma^{Frag}_{LC}} {d^3\vec{P}_h} &= \frac{2M_h\alpha_{em}^2} {S}\,\epsilon_\perp^{P_{h\perp}S_{hT}}\sum_q e_q^2\int_{z_{min}}^1\!\frac{dz} {xz^3}\frac{1} {S+T/z}\,\frac{1} {-\hat{t}-x\hat{u}}\,f_1^q(x)\nonumber\\
&\times\left[z\frac{d\hat{D}_T^q(z)} {dz}\,\hat{\sigma}_D + \hat{D}_T^q(z)\,\hat{\sigma}_N + \frac{1} {z}D_T^q(z)\,\hat{\sigma}_2 + \int_z^\infty\!\frac{dz_1} {z_1^2} \,\frac{1} {1/z-1/z_1}\frac{1} {\xi}\,\hat{D}_{FT}^{q,\Im}(z,z_1)\,\hat{\sigma}_3\right]\,, \label{e:sigmaFLC}
\end{align}
where 
\begin{align}
\hat{\sigma}_D = (1-x)\,\hat{\sigma}_U\,,\quad\quad\quad\quad\quad\hat{\sigma}_N &= -\frac{x\hat{s}} {\hat{t}}\,\hat{\sigma}_U\,,\\[0.1cm]
\hat{\sigma}_2 = \frac{2\hat{s}((x-2)(\hat{s}-\hat{u})\hat{t}-2x\hat{s}\hat{u})} {\hat{t}^3}\,,\quad\quad \hat{\sigma}_3&= -\frac{2xz^2\hat{s}} {\hat{t}}\,\hat{\sigma}_U\,, 
\end{align}
with $\hat{\sigma}_U = 2(\hat{s}^2+\hat{u}^2)/\hat{t}^2$ the hard part for the unpolarized cross section.

We now turn to the calculation of the distribution term, where we follow similar steps to the fragmentation case (see also, e.g., Ref.~\cite{Zhou:2010ui}).  From the diagram in Fig.~\ref{f:graphs}(a), we keep the transverse momentum $p_\perp$ of the active quark, leading to a derivative and non-derivative piece involving the first $p_\perp$-moment of the Boer-Mulders function.  However, we will not get any contribution from neglecting $p_\perp$ and using a twist-3 Dirac projection because the associated function $h^q(x)$ vanishes due to time-reversal invariance~\cite{Goeke:2005hb}.\footnote{In principle, one could have a contribution from $e^q(x)$, but its hard factor vanishes for $A_N$ (but it would be nonzero for $A_{LT}$).}  Thus, we have for the quark-quark correlator term,
\begin{align}
\frac{P_h^0d\sigma^{Dist}_{LC,qq}} {d^3\vec{P}_h} &= \frac{8\pi M\alpha_{em}^2} {S}\,\epsilon_\perp^{P_{h\perp}S_{hT}}\sum_q e_q^2\int_{z_{min}}^1\!\frac{dz} {xz^3}\frac{1} {S+T/z}\,\frac{1} {\hat{u}}\,H_1^q(z)\nonumber\\
&\hspace{4cm}\times\left\{x\,\frac{dH_{FU}^q(x,x)} {dx}\!\left[\frac{\hat{s}\hat{u}} {\hat{t}^2}\right] - H_{FU}^q(x,x)\!\left[\frac{\hat{s}\hat{u}(\hat{u}-\hat{s})} {\hat{t}^3}\right]\right\}, \label{e:sigma_dist_qq}
\end{align}
where we have used the identity in Eq.~(\ref{e:HFU_BM}), and $d\sigma^{Dist}_{LC}$ indicates the distribution term in lightcone gauge.

Next, we look at the quark-gluon-quark correlator term from the graph in Fig.~\ref{f:graphs}(b).  Like in the fragmentation case, we initially have a matrix element involving $A_\perp$ that we rewrite in terms of $F_\perp^{n\alpha}$ (see Eq.~(\ref{e:HFU})), which generates a factor $1/(x-x_1)$.  The advanced BC dictates that we regulate the pole at $x=x_1$ by $1/(x-x_1)\to 1/(x-x_1-i\epsilon)$.  Typically in lightcone gauge one would neglect the transverse momentum $p_\perp$, $p_{1\perp}$ of the quarks.  However, since the quark propagator in the hard part is
\begin{equation}
\frac{i(\slashed{p}_1+\slashed{q})}{(p_1+q)^2+i\epsilon}=\frac{i(z/T)(\slashed{p}_1+\slashed{q})} {x-x_1+x_{1\perp}-i\epsilon}\,, \label{e:prop}
\end{equation}
where $x_{1\perp}$ is a scalar that depends on $p_{1\perp}$, doing so would create a divergence when one picks out the pole at $x_1 = x$.  That is, one would have the factor
\begin{equation}
\frac{1} {x-x_1-i\epsilon}\frac{1} {x-x_1+x_{1\perp}-i\epsilon} \overset{x_{1\perp}\!\to0} {=} \frac{1} {x-x_1-i\epsilon}\frac{1} {x-x_1-i\epsilon}\,,
\end{equation}
which would lead to 1/0 when summing over residues.  So in a first step we keep the transverse momenta of the quarks.  One can then carry out the $x_1$-integral in the cross section as follows (for brevity, we only keep the $x,x_1$ dependence in the arguments):
\begin{align}
&\int dx_1\,\frac{1} {x-x_1-i\epsilon}\frac{1} {x-x_1+x_{1\perp}-i\epsilon}\,H_{FU}^q(x,x_1)\hat{S}_{(b)}(x,x_1) \nonumber\\
&= -i\pi\, \frac{1} {x_{1\perp}}\left\{\!\left[H^q_{FU}(x,x+x_{1\perp})\hat{S}_{(b)}(x,x+x_{1\perp})-H^q_{FU}(x,x)\hat{S}_{(b)}(x,x)\right]\!\right\}\nonumber\\
&\left.=-i\pi\frac{\partial} {\partial x_1}\left[H^q_{FU}(x,x_1)\hat{S}_{(b)}(x,x_1)\right]\right|_{x_1\,= \,x }, \label{e:x1int}
\end{align}
where $\hat{S}_{(b)}(x,x_1)$ represents the entire hard factor in Fig.~\ref{f:graphs}(b), and the last line holds in the limit $p_\perp,p_{1\perp}\!\to \!0$.  Note that the simplification in the last line of (\ref{e:x1int}) is a direct consequence of choosing the advanced BC.  From here, one can calculate the quark-gluon-quark correlator term, and finds
\begin{align}
\frac{P_h^0d\sigma^{Dist}_{LC,qgq}} {d^3\vec{P}_h} &= \frac{8\pi M\alpha_{em}^2} {S}\,\epsilon_\perp^{P_{h\perp}S_{hT}}\sum_q e_q^2\int_{z_{min}}^1\!\frac{dz} {xz^3}\frac{1} {S+T/z}\,\frac{1} {\hat{u}}\,H_1^q(z)\nonumber\\
&\hspace{4cm}\times\Bigg\{x\,\frac{dH^q_{FU}(x,x)} {dx}\left[\frac{\hat{s}\hat{u}^2} {\hat{t}^3}\right] + H^q_{FU}(x,x)\left[\frac{\hat{s}\hat{u}(\hat{u}-\hat{s})} {\hat{t}^3}\right]\Bigg\}\,. \label{e:sigma_dist_qgq}
\end{align}
We mention that in deriving (\ref{e:sigma_dist_qgq}) we have also made use of the relation 
\begin{equation}
\left.\frac{\partial} {\partial x_1} H^q_{FU}(x,x_1)\right|_{x=x_1} = \frac{1} {2} \frac{d} {dx} H^q_{FU}(x,x)\,,
\end{equation}
which follows from the symmetry property of the correlator:~$H^q_{FU}(x,x_1) = H^q_{FU}(x_1,x)$.  Combining Eqs.~(\ref{e:sigma_dist_qq}), (\ref{e:sigma_dist_qgq}) gives
\begin{align}
\frac{P_h^0d\sigma^{Dist}_{LC}} {d^3\vec{P}_h} &= \frac{8\pi M\alpha_{em}^2} {S}\,\epsilon_\perp^{P_{h\perp}S_{hT}}\sum_q e_q^2\int_{z_{min}}^1\!\frac{dz} {xz^3}\frac{1} {S+T/z}\,\frac{1} {\hat{u}}\,H_1^q(z)\left(x\,\frac{dH^q_{FU}(x,x)} {dx}\right)\left[-\frac{\hat{s}^2\hat{u}} {\hat{t}^3}\right]. \label{e:sigma_dist}
\end{align}
Note that the hard parts for the non-derivative term cancel between the $qq$ and $qgq$ diagrams, leaving us with a final result that only involves a derivative piece, which is consistent with what is stated in Sec.~III of Ref.~\cite{Zhou:2008fb}.  This derivative term also agrees with the $qq^\prime\to qq^\prime$ channel result in \cite{Kanazawa:2000cx}.  The formulas in Eqs.~(\ref{e:sigmaFLC}), (\ref{e:sigma_dist}) (and their verification in the Feynman gauge calculation of Sec.~\ref{s:Feyn}) are the main results of this paper.\\

%
%
\section{Feynman gauge calculation} \label{s:Feyn}

Here we present some details of our calculation in Feynman
gauge. The collinear twist-3 formalism in Feynman gauge has been 
widely studied in the
literature~\cite{Qiu:1991pp,Qiu:1998ia,Kouvaris:2006zy,Eguchi:2006mc,Zhou:2009jm,Kanazawa:2013uia}
and well-established both for pole and non-pole contributions at
LO in QCD perturbation theory. In
the following we summarize some of the key steps to derive the single-spin dependent
cross section for the distribution and fragmentation terms in $\ell\, p\to\Lambda^\uparrow X$ based on~\cite{Eguchi:2006mc,Kanazawa:2013uia}.

As in the case of the lightcone gauge calculation, we consider the twist-3
contribution from the diagrams in Fig.~\ref{f:graphs}. 
A standard and
systematic method to extract those effects is the collinear expansion of the
hard parts $\Sh_{(a)}$, $\Sh_{(b),
L(R)}^\alpha$, and $\Sh_{(c), L(R)}^\alpha$. 
Here the
subscript $L \,(R)$ tells us that the 
coherent gluon resides on the left-side (right-side) of the final-state
cut, and the color indices for $\hat{S}_{(b)}$, $\hat{S}_{(c)}$ have been suppressed for
simplicity. 
Performing the collinear expansion and recasting the field operators, it
is easy to see that the twist-3 cross section can be expressed in terms
of the gauge invariant F-type matrix elements, defined in
Sec.~2, and several terms that contain gauge noninvariant matrix elements.
Our task is to show the latter ones either
vanish or combine into other gauge invariant operators at ${\cal O}(g)$ 
to ensure that the cross section contains only gauge invariant
matrix elements at twist-3 accuracy. 
To this end, we first need to find particular relations
among the hard parts $\Sh_{(a,b,c)}$.
A convenient way to achive this is to make use of
Ward-Takahashi identities (WTI) in QCD.
The WTI for the present process can
be obtained by contracting the Lorentz index of the hard part for the
$qgq$ diagrams in Figs.~\ref{f:graphs}(b), (c) with the
scalar-polarized coherent gluon as
\begin{align}
&(k-k_1)_\alpha\, \hat{S}_{(c),L}^{a,\alpha} (x,k_1,k) \vb_{\rm non-pole} \;\;\;=
  T^a \hat{S}_{(a)}(x,k)\,, \label{nonp}\\
 &\,(p-p_1)_\alpha\, \hat{S}_{(b),L}^\alpha (p_1,p,z) \vb_{\rm pole} \hspace{1cm}\,=
 0\,, \label{pole}
\end{align}
and similarly for $\Sh_{(b),R}^\alpha$ and $\Sh_{(c),R}^\alpha$. 
Note that the above relations are the ones before taking the collinear
limit and in practice one needs the expression of these and their first
derivatives in that limit. 
For the non-pole (fragmentation) contribution,
using such identities derived from Eq.~(\ref{nonp}) allows us to
reorganize all the operators into a manifestly
color gauge invariant form without explicitly calculating the Dirac
or color factors. 
On the other hand, for the pole (distribution) piece the
remaining gauge noninvariant terms vanish, and,
thus, the twist-3 cross section is expressed solely in terms of the F-type
function.
After some algebra we obtain a gauge invariant
expression of the single-spin dependent cross section as
\begin{eqnarray}
 \frac{P_h^0 d\sigma_{Feyn}^{Frag}}{d^3\vec{P_h}} &\!\!\!=\!\!\!& 
  \frac{M_h \alpha_{em}^2}{S} \sum_{q}
  e_q^2 \int\! \frac{dx}{x} f_1^q(x) \nn\\
&& \times\, \Bigg\{ \epsilon_\perp^{\alpha S_{hT}} \int\! \frac{dz}{z^2} \, D^q_T(z) 
   \, \Tr{ \gamma_\alpha \, \Sh_{(a)}(x,z)} -
   \frac{\epsilon_\perp^{\alpha S_{hT}}}{2} \int \frac{dz}{z^2} \,
   \hat{D}^q_T(z) \,
 \Tr{ \slashed{P_h} \,
  \frac{\partial \Sh_{(a)}(x,k)}{\partial k^\alpha} \cl } \nn\\[0.3cm]
&&  \hspace{-0.45cm}+ \,2 \int\! \frac{dz_1}{z_1^2} \int\!\frac{dz}{z}
 PV{\frac{1}{1/z-1/z_1}} \nn\\
 && \times\, \Tr{ \left( \slashed{P_h}
				\epsilon_\perp^{\alpha S_{hT}}
				\hat{D}_{FT}^{q,\Im} (z,z_1) + i \gamma_5
				\slashed{P_h} S_{hT}^\alpha \hat{G}_{FT}^{q,\Im}
				(z,z_1)\right)
\, \Sh_{(c),L,\alpha}(x,z_1,z) } \Bigg\}\,, \\[0.3cm]
 \frac{P_h^0 d\sigma_{Feyn}^{Dist}}{d^3\vec{P_h}} &\!\!\!=\!\!\!& \frac{M
 \alpha_{em}^2}{S}  \left( \frac{i\epsilon_\perp^{\alpha\beta}}{2}
		    \right) \sum_{q}
  e_q^2 \int\! \frac{dz}{z^2} H_1^q(z) \int \!dx_1 \int\! dx 
  \,H^q_{FU}(x,x_1) \nn\\
 &&\times\, \Tr{ \gamma_\beta \slashed{P} \gamma_5
  \frac{\partial \Sh_{(b),\sigma} (p_1,p,z)
  P^\sigma}{\partial p^\alpha} \cl },
\end{eqnarray}
where $\Sh_{(b),\sigma} (p_1,p,z) \equiv \Sh_{(b),L,\sigma} (p_1,p,z) +
\Sh_{(b),R,\sigma} (p_1,p,z)$ and ``c.l.'' denotes the collinear
limit $p^\alpha \to xP^\alpha$ and $k^\alpha\to P_h^\alpha/z$. Note in
the above formula we have reproduced the gauge invariant twist-3 PDFs and FFs
defined in Sec.~2 at
${\cal O}(g)$ including the Wilson line and its derivative. 
Computing the trace and applying the relation (\ref{e:useful}), one
eventually finds agreement with Eqs.~(\ref{e:sigmaFLC}),
(\ref{e:sigma_dist}). Thus, we have demonstrated that lightcone gauge and Feynman gauge lead to identical results for both the twist-3 distribution
and fragmentation terms in $\ell\,p\to\Lambda^\uparrow X$.  In addition, we point out that one can obtain the single-spin dependent cross section for $\ell \,p^\uparrow \to \pi\,X$ from our work here through a straightforward replacement of Dirac projections in the correlators.  We performed this task and verify that the results in~\cite{Gamberg:2014eia} are the same in both gauges. 
%
%
\section{Summary and conclusion} \label{s:sum}
In this paper we have analyzed the transverse SSA in $\ell\,p\to\Lambda^\uparrow X$, calculating both the distribution and fragmentation terms.  The computation of the latter is the first time this piece has been derived for transversely polarized hadrons within collinear twist-3 factorization. We found the same result in both lightcone gauge and Feynman gauge, and also verified that the distribution and fragmentation terms in $\ell\,p^\uparrow\to \pi\,X$ (that appear in~\cite{Gamberg:2014eia}) are gauge invariant.  For the lightcone gauge calculation of the fragmentation term, we followed the procedure in Ref.~\cite{Metz:2012ct} for $p^\uparrow p\to \pi\,X$, where the fragmentation mechanism has the potential to be extremely important to that reaction~\cite{Kanazawa:2014dca} (see also~\cite{Lu:2015wja}).  Therefore, it is encouraging that using the techniques of~\cite{Metz:2012ct} provide a gauge invariant result for the fragmentation part of $\ell\,p\to\Lambda^\uparrow X$ and $\ell\,p^\uparrow\to \pi\,X$.  Moreover, the leptoproduction of transversely polarized $\Lambda$ hyperons could be another interesting observable to test the origin of transverse SSAs, and because of this possibility we plan to supplement this analytical work with a numerical study in the future.  We will then address the recently obtained data from the HERMES Collaboration~\cite{Airapetian:2014tyc} and explore potential measurements at facilities like a next generation Electron-Ion Collider~\cite{Boer:2011fh,Accardi:2012qut}.

\section*{Acknowledgments}

We would like to thank J.~Zhou for a useful discussion about lightcone gauge calculations.  This work has been supported by the National Science
Foundation under Contract No.~PHY-1205942 (K.K. and A.M.), and the RIKEN BNL
Research Center (D.P.).

\end{document}